\documentclass[1p]{elsarticle}

\usepackage{lineno,hyperref}
\usepackage{mathtools}
\usepackage{algorithm,algpseudocode}
\modulolinenumbers[1]
\algdef{SE}[SUBALG]{Indent}{EndIndent}{}{\algorithmicend\ }%
\algtext*{Indent}
\algtext*{EndIndent}

\journal{Knowledge-Based Systems}

\begin{document}

\begin{frontmatter}

\title{Electoral Forecasting Using a Novel Temporal Attenuation Model: Predicting the US Presidential Elections}

\author{Alexandru Top\^{i}rceanu}
\address{Department of Computer and Information Technology, Politehnica University Timi\c{s}oara, Timi\c{s}oara, Romania}
\fntext[myfootnote]{Corresponding author email: alext@cs.upt.ro}




\begin{abstract}
Electoral forecasting is an ongoing scientific challenge with high social impact, as current data-driven methods try to efficiently combine statistics with economic indices and machine learning. However, recent studies in network science pinpoint towards the importance of temporal characteristics in the diffusion of opinion.
As such, we combine concepts of micro-scale opinion dynamics and temporal epidemics, and develop a novel macro-scale temporal attenuation (TA) model, which uses pre-election poll data to improve forecasting accuracy. 
Our hypothesis is that the timing of publicizing opinion polls plays a significant role in how opinion oscillates, especially right before elections.
Thus, we define the \textit{momentum} of opinion as a temporal function which bounces up when opinion is injected in a system of voters, and dampens down during states of relaxation. We validate TA on survey data from the US Presidential Elections between 1968--2016, and TA outperforms statistical methods, as well the best pollsters at their time, in 10 out of 13 presidential elections. We present two different implementations of the TA model, which accumulate an average forecasting error of 2.8--3.28 points over the 48-year period. Conversely, statistical methods accumulate 7.48 points error, and the best pollsters accumulate 3.64 points. Overall, TA offers increases of 23-37\% in forecasting performance compared to the state of the art.
We show that the effectiveness of TA does not drop when relatively few polls are available; moreover, with increasing availability of pre-election surveys, we believe that our TA model will become a reference alongside other modern election forecasting techniques.

\end{abstract}


\end{frontmatter}


\section{Introduction}

Understanding the dynamics of information, which shapes many aspects of our social lives, is a major research drive in our increasingly networked society \cite{granovetter1973strength,barabasi2002linked,lazer2009life,borner2018forecasting}. 
Be it under the form of a commercial, a rumour, a virus, or a blog post, information diffusion (or propagation) receives substantial attention from multidisciplinary fields of research \cite{pastor2015epidemic,jackson2002evolution,easley2010networks}. Out of these, the ability to predict election outcomes is just one of many areas of research that sees benefits from cutting-edge investigation techniques, like social media analysis and data science \cite{golbeck2013method,rodriguez2011uncovering,guille2012predictive}. 

Research on forecasting election polls was originally constructed by employing classic statistical models, applied on opinion polls prior to the election day \cite{nadeau2010electoral,whiteley1979electoral}. Ever since the late '70s, it became a scientific fact that correct timing of the election date can be crucial for the outcome \cite{whiteley1979electoral}.
Election forecasting -- especially in the case of presidential elections --  employs so-called macromodels \cite{jensen2017winners}; these are statistical models based on national economic and political fluctuations. On the other hand, micromodels are models based on surveys of individual voters during the pre-election period \cite{lewis1999voters}. 
Current state of the art in forecasting employs multilevel regression and post-stratification \cite{christensen2008predicting,kiewiet2018predicting,hummel2014fundamental} (MRP). Hence, we mention several reputable institutions in the United States, dedicated to the analysis of election data using variants of MRP, like \textit{Real Clear Politics}, \textit{Huffington Post}, \textit{FiveThirtyEight}, \textit{Daily Kos}, or \textit{Understanding America Study}.
Arching over the specific methodologies employed by these platforms, we summarize them as follows: 
\begin{itemize}
\item Poll weighing/averaging -- polls from different sources are weighted based on the pollster credibility.
\item Poll adjustment -- number of likely voters, convention influence, omission of third party candidates are taken into account.
\item Adding demographic and optional economic data -- used to scale surveys at state and national levels.
\item Simulating -- using a probabilistic distribution to account for uncertainty in the data.
\end{itemize}

Apart from techniques like MRP, we find studies proving that alternative subjective surveying methods may also be efficient forecasters. Thus, the American National Election Surveys from 1956 to 1996 show that voters could themselves better forecast who will win the presidential elections \cite{lewis1999voters}. Another study shows that quick and unreflective facial judgments of gubernatorial candidates are more accurate in predicting the winners than deliberating on the competence of each candidate \cite{ballew2007predicting}.
In essence, voter forecasting models derived from vote expectations, represent a promising alternative to classic statistical approaches \cite{lewis1989citizen}.

Bridging over to social networks and media, there is a scientific debate on how the wide coverage of publicized opinion polls in media can affect voters before election \cite{weimann1990obsession}. 
It is already known that social networks have a decisive role in the diffusion of information, and have proven to be very powerful in many situations involving macroscopic behavior \cite{golbeck2013analyzing,golbeck2013method}. Examples include, decisively influencing the Arab Spring in 2010 \cite{hussain2013best,howard2011opening}, and the U.S. presidential elections in 2008
\cite{hughes2009twitter}, and 2012 \cite{conway2013twitter}. 
Analyzing the dynamics of this social layer can offer substantial predictive power over the real-world social networks they model. Studies on diffusion predictability are further found in marketing and public relations \cite{easley2010networks,papasolomou2012social},
epidemic spreading\cite{hufnagel2004forecast}, hurricane forecasting \cite{gladwin2007social},
or forecasting box-office revenues of movies \cite{asur2010predicting}.

Network science proposes understanding diffusion processes by designing interactions at micro-scale level \textit{(i.e.}, between individual social agents), and forecasting opinion evolution at macro-scale level \cite{jackson2017economic}. Specifically, the macroscopic behavior is inferred by: (i) monitoring when social agents become \textit{indoctrinated} by their neighborhood (\textit{i.e.}, they adopt information, get infected, buy merchandise \cite{pastor2015epidemic,easley2010networks,barabasi2016network}), then, (ii) predicting how cascades of information flow, and how the diffusion process is percolated by individuals. 
Nevertheless, temporal aspects are shown to play an essential role in the diffusion of influence \cite{rodriguez2011uncovering,guille2012predictive}, as does the timing of publicizing opinion polls \cite{weimann1990obsession} and even organizing elections \cite{whiteley1979electoral}. 

Consequently, this paper builds upon the premises that we are able to extrapolate the macroscopic behavior of a society (here, in the context of elections) by inferring microscopic temporal dynamic models during the pre-election period. Thus, our contributions in this paper are:

\begin{itemize}
\item We formulate an analytic methodology for modeling the macro-scale evolution of a multi-opinion system, targeting better election poll forecasting.
\item We define an experimental setup, based on pre-election data, to validate the underlying assumptions of our approach.
\item We present a comprehensive case study on US Presidential Elections to measure the efficiency of our approach against state of the art forecasting estimates, including MRP, as recently used by the best pollsters in the USA.
\item We explore the feasibility of applying TA in real time, during an ongoing pre-election period, and compare its performance to MRP at several points in time, relative to the election day.
\end{itemize}


\section{The temporal attenuation model}

Application of TA starts by gathering a set of pre-election multi-opinion polls with temporal information, \textit{i.e.}, corresponding to specific relative dates before an election. Based on the number of candidates for election, we define temporal poll vectors $p_i(t)$, where $i$ is the index of the candidate in the multi-opinion system, and $t$ is the time (date) of the poll. 
We define as \textit{opinion injection}, at any time $0 \leq t < d$, all discrete observations which stem from the public opinion polls $p_i(t)$, preceding the election day $d$.
We further define a discrete temporal election axis $t=[0,d)$ as being relative to the date of the first opinion poll (which becomes $p_i(t=0)$), and the last opinion poll prior to the election day $t=d$. 

An essential aspect of TA is to reproduce the real-world timing when injecting opinion, \textit{i.e.}, in this paper, at the level of day, as explained in Figure \ref{fig:ta-model}a. As such, we do not condense consecutive polls, one after another, but scatter them along the temporal axis in order to model bounces and periods of relaxation mirroring the real-world general opinion.

Given the available poll vectors $p_i(t)$, we construct dataset $P$ consisting of daily opinion corresponding to each candidate. Since there may be days without available polls, we compensate every such day $0<k<d$ by adding a 0 (no vote) for each candidate. Consequently, we can describe dataset $P$ and an individual poll vector $p_i(t)$ as: 

\begin{equation}
P=\{p_i(t=0),p_i(t=1),...,p_i(t<d)\}
\label{eq:P1}
\end{equation}

\begin{equation}
p_i(t)=
\begin{cases}
\{\frac{p_0^*(t)}{\sum p_i^*(t)},..., \frac{p_n^*(t)}{\sum p_i^*(t)} \}, \;\; \mbox{if $\exists$ poll for $t$} \\
\{0,...,0\}, \;\; \mbox{otherwise}
\end{cases}
\label{eq:P2}
\end{equation}

Equation \ref{eq:P1} specifies that $P$ consists of continuous (daily) poll vectors $0 \leq t < d$. Equation \ref{eq:P2} specifies that if a poll is available for a specific day, then we calculate a normalized opinion value based on the raw poll data $p^*$ (\textit{e.g.}, number of voters, vote percentages).
In the validation data we have polls ranging from a few hundred to tens of thousands of voters, so that a normalization of the raw amplitudes is recommended.

\subsection{Micro-scale interactions}

Most existing micro-scale models for opinion injection rely on fixed thresholds,  or thresholds evolving according to simple probabilistic processes \cite{axelrod1997dissemination,goldenberg2001talk,topirceanu2016tolerance}, and are impervious to any temporal aspects \cite{guille2013information}.
Nonetheless, studies applied in epidemiology propose three popular parametric models for the likelihood of disease transmission rates, considering time as a parameter \cite{wallinga2004different,myers2010convexity}: power-law, exponential, and Rayleigh. These functions model the temporal damping (fading) of infectiousness after exposure, yet, they may be used to trace the evolution of opinion after each injection.

In this paper, we introduce the power-law (PTA) and exponential (ETA) temporal attenuated models, and analyze their efficiency. In their continuous epidemiological formulation \cite{myers2010convexity}, these models express the transmission likelihood $\lambda_i(t)$ of a disease in time, after a relative time $\Delta t$ since an individual $i$ was infected, as expressed by the following expression: $\lambda_i(t) = \alpha_i \cdot \Delta t^{-\beta_i}$ for PTA, respectively $\lambda_i(t) = \alpha_i \cdot e^{-\Delta t \beta_i}$ for ETA.

Additionally, we parameterize the two TA models with an amplitude factor $\alpha_i$ and a damping factor $\beta_i$, specific for every candidate $i$. The $\alpha_i$ factor determines the amplitude of the positive bounce when opinion is injected, and the $\beta_i$ factor controls the damping speed towards the relaxed state ($\lambda(t\rightarrow\infty)=0$) for any candidate (see Figure \ref{fig:ta-model}b).

\subsubsection{Macro-scale emergent behavior}

Based on the introduced temporal micro-scale models, we define the concept of opinion momentum $M_i(t)$. The momentum of each candidate $i$ is an aggregated macro-scale estimator for the evolution of opinion of the entire voter system. 
We extrapolate $M_i(t)$ for PTA and ETA as follows:

\begin{equation}
M_i(t) =
\begin{cases}
\alpha_i(t) \cdot t^{-\beta_i} \;\;\; (PTA)\\
\alpha_i(t) \cdot e^{-t \beta_i} \;\; (ETA)
\end{cases}
\label{eq:momentum}
\end{equation}

The amplitude $\alpha_i$ evolves according to the following rules: if we are during a relaxation state, when there is no opinion injection at moment $t$ (i.e., $p_i(t)=0$), then $\alpha_i$ remains unchanged. Consequently, as we progress on the discrete time axis ($t \rightarrow t+1$), momentum $M_i(t)$ will decrease. On the other hand, if opinion is injected and we have a poll $p_i(t)>0$ at the current moment, then $\alpha_i$ is increased by an amplitude proportional to the normalized number of votes $p_i(t)$. The evolution in time of $\alpha_i$ is given by the following equation:

\begin{equation}
\alpha_i(t) = 
\begin{cases}
\alpha_i(t-1) t^{-\beta_i} + p_i(t), & \text{if }\;\,p_i(t)>0\\
\alpha_i(t-1), & \text{if }\;p_i(t)=0
\end{cases}
\label{eq:alpha}
\end{equation}

By simulating the evolution of each opinion momentum in time, we can infer the current opinion $\Omega_i$ by normalizing the momentums of each candidate $i$ as follows:

\begin{equation}
\Omega_i(t) = M_i(t) / \sum_{j} M_j(t)
\label{eq:omega}
\end{equation}

The process of evolving momentums $M_i(t)$ and opinions $\Omega_i(t)$ is detailed on a proof of concept voting system in Figure \ref{fig:ta-model}. A detailed comparative analysis of TA is further provided given in \textit{Appendix A}. 

\begin{figure}[!htb]
\hspace*{-3cm}
\includegraphics[width=1.4\linewidth]{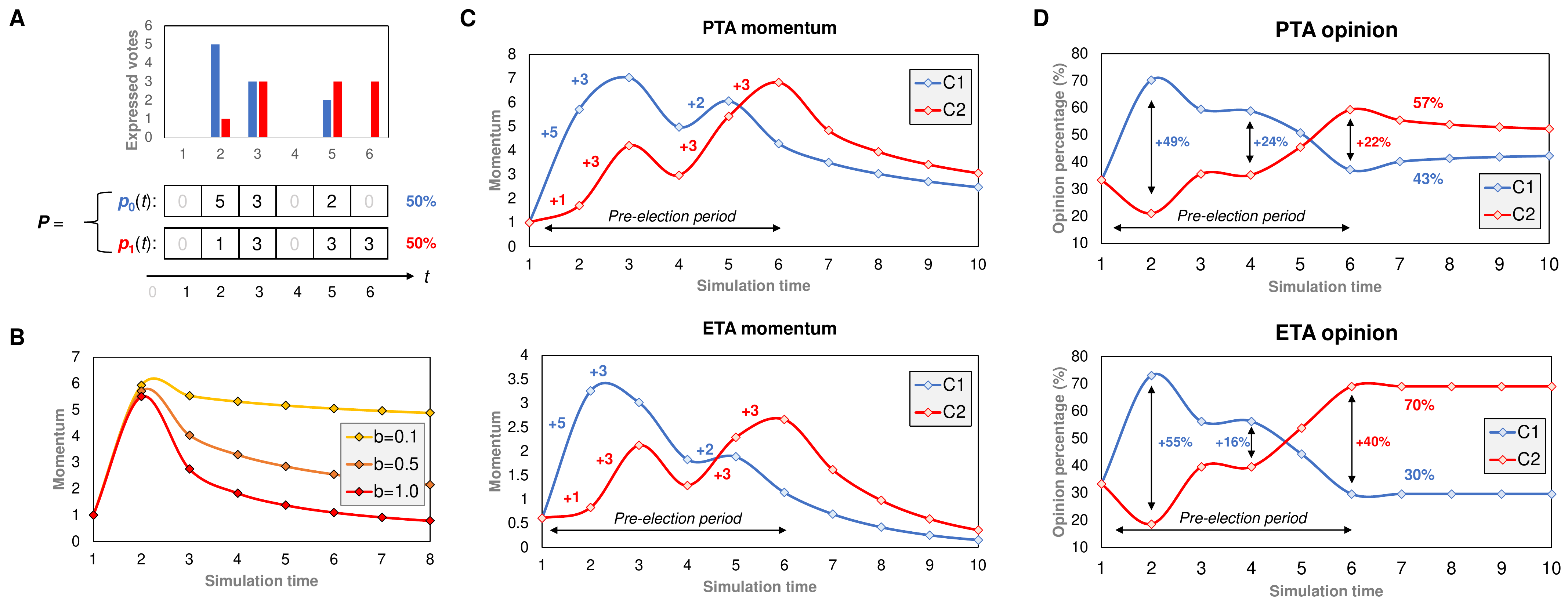}
\caption{Overview of the temporal attenuation (TA) model applied on two candidates (C1-blue and C2-red) receiving an equal number of votes (\textit{i.e.}, suggesting a 50-50\% tie) over 6 days before election, but according to different temporal patterns. \textit{(A)} Surveys are collected for the two candidates at $t=\{2,3,5,6\}$. From these, set $P$ is assembled consisting of poll vectors $p_0(t)$ and $p_1{t}$. \textit{(B)} Impact of damping factor $\beta$ on a TA function with one spike at $t=2$, followed by a continuous relaxation state. A higher $\beta$ translates into a more abrupt damping of the momentum. \textit{(C)} Momentum $M_i(t)$ evolution for PTA and ETA corresponding to the poll vectors $p_0(t)$ and $p_1(t)$. Individual votes are displayed in absolute value on the graphs. The simulation using dataset $P$ corresponds to the pre-election period ($0<t\leq 6$). \textit{(D)} Opinion $\Omega_i(t)$ evolution for PTA and ETA corresponding to the momemtums in panel (c). Several poll differences are displayed at $t=\{2,4,6\}$ using the color of the virtual winner at that moment.}
\label{fig:ta-model}
\end{figure}

\subsection{Temporal attenuation algorithm}

\begin{figure}[!htb]
\includegraphics[width=1\linewidth]{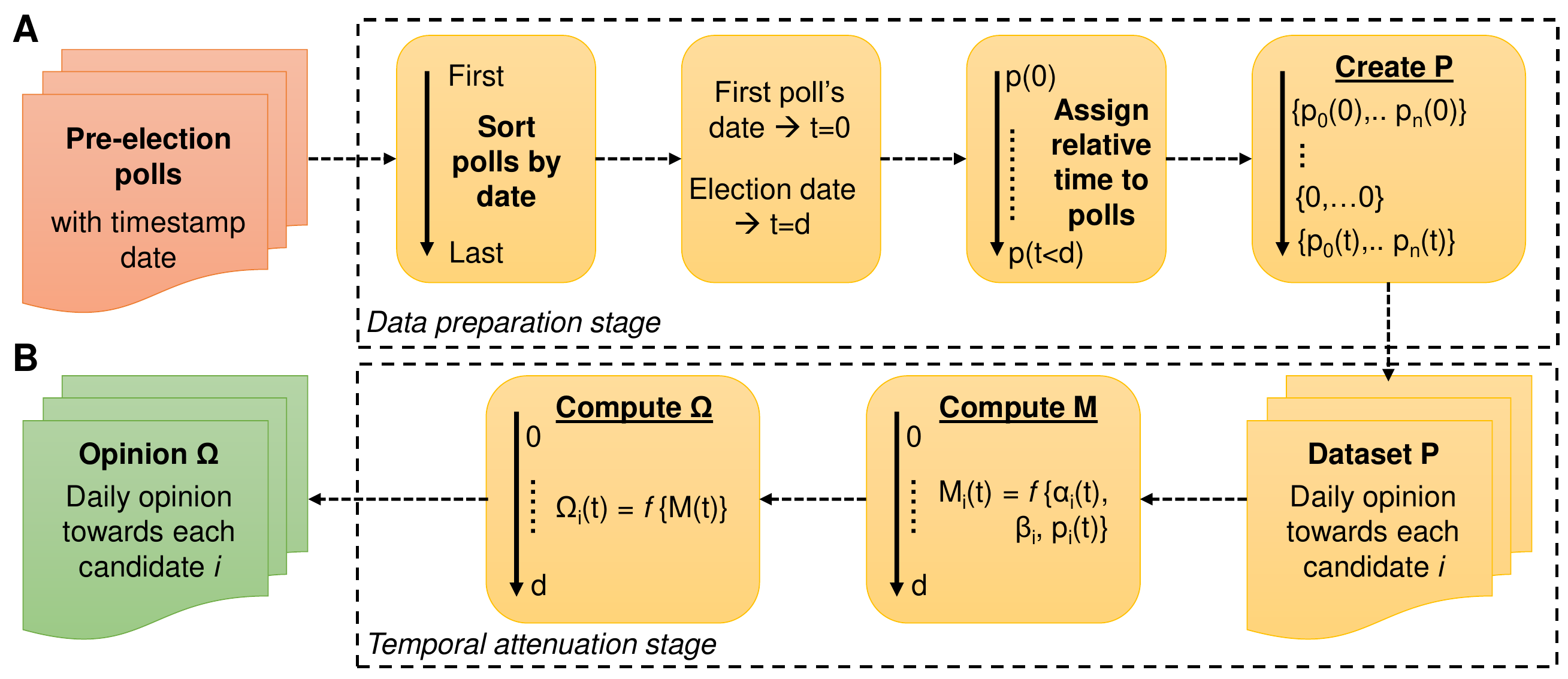}
\caption{Flowchart of applying TA. \textit{(A)} The data preparation stage takes as input a set of pre-election polls with timestamp, and produces the daily opinion set of poll vectors $P$ as an intermediary output. Polls are first sorted by date from first to last, then the first date is assigned as relative day $t=0$, while the election day is assigned as relative day $t=d$. Each other poll gets assigned a relative day $0\leq t < d$. $P$ is created by either setting a normalized poll for that day ($p_i(t)$), or, if no poll is available, an empty poll vector $\{0,...0\}$. \textit{(B)} The temporal attenuation stage further takes $P$ as input to compute the momemtum $M(t)$, based on  $\alpha_i(t)$, $\beta_i$, and $p_i(t)$. Consequently, the daily opinion $\Omega(t)$ towards each candidate is computed as output from $M(t)$.}
\label{fig:ta-algorithm}
\end{figure}

By corroborating all the introduced terms, we present the flowchart of applying TA on a pre-election dataset in Figure \ref{fig:ta-algorithm}, and provide the supporting algorithmic pseudocode in Algorithm \ref{alg:ta-algo}. The required input is a dataset consisting of pre-election polls, where each poll expresses opinion for each candidate $i$ in the multi-opinion system. The first stage of the algorithm (\textit{data preparation}), depicted in Figure \ref{fig:ta-algorithm}a, creates the intermediary dataset $P$ consisting of daily poll vectors $p_i(t)$ for each day $0 \leq t < d$. The second stage of the algorithm (\textit{temporal attenuation}), depicted in Figure \ref{fig:ta-algorithm}b, computes the momentum of opinion $M_i(t)$ from $P$ using the amplitude $\alpha_i(t)$, damping factor $\beta_i$, and daily poll vectors $p_i(t)$. The output, represented by the daily opinion evolution $\Omega_i(t)$, is computed from each momentum $M(t)$.

\begin{algorithm}
\caption{Electoral forecasting algorithm based on temporal attenuation (TA)}\label{alg:ta-algo}
\begin{algorithmic}[1]
\State \textbf{Input:} Pre-election polls for each candidate $i$, with timestamp
\State \textbf{Stage A:} \textit{Data preparation}
\Indent
\State \textbf{sort} polls by date in increasing order
\State \textbf{assign} first poll's date as day $t \leftarrow 0$
\State \textbf{assign} election date as day $t \leftarrow d$
\State \textbf{assign} relative day $0 \leq t < d$ for $\forall$ pre-election poll $\rightarrow p_i(t)$
\State \textbf{for} each day $t \in [0,d)$ assign $p_i(t)$ according to Equation \ref{eq:P2} $\rightarrow P$
\State \textbf{output:} $P=\{p_i(t=0),p_i(t=1),...,p_i(t<d)\}$ 
\EndIndent
\State \textbf{Stage B:} \textit{Temporal attenuation}
\Indent
\State \textbf{input:} $P$ (Equation \ref{eq:P1})
\State \textbf{compute} opinion momentum $M_i(t)$:
\For{$\forall$ candidate $i$ in the multi-opinion system $P$}
\State $\alpha_i(0) \leftarrow p_i(0)$
\For{$t \in [1,d)$}
\If{$p_i(t)>0$} (Equation \ref{eq:alpha})
\State $\alpha_i(t) \leftarrow \alpha_i(t-1)\cdot t^{-\beta_i} + p_i(t)$
\Else
\State $\alpha_i(t) \leftarrow \alpha_i(t-1)$
\EndIf
\State $M_i(t) \leftarrow \alpha_i(t)\cdot t^{-\beta}$  (Equation \ref{eq:momentum})
\EndFor
\EndFor
\State \textbf{compute} opinion $\Omega_i(t)$:
\For{$t \in [0,d)$}
\State $\Omega_i(t) = M_i(t) / \sum_{j} M_j(t)$ (Equation \ref{eq:omega})
\EndFor
\EndIndent
\State \textbf{Output:} Evolution of daily opinion $\Omega_i(t)$ towards each candidate $i$ in time $0 \leq t < d$
\end{algorithmic}
\end{algorithm}

\subsection{Experimental validation}

Prediction models, where microscopic interactions are considered, consistently represent data as information cascades triggered by so-called opinion sources (also spreader nodes, stubborn agents, vital nodes) \cite{axelrod1997dissemination,goldenberg2001talk,topirceanu2016tolerance}.
While these interactions are meaningful, accounting for all of them is still technologically, ethically, and legally impossible \textit{(e.g.}, analyzing \textit{all} tweets posted by \textit{all} users in the USA before an election). As such, we use ubiquitous data under the form of pre-election opinion polls, which are centralized and open, gathered from \textit{Real Clear Politics}, \textit{Understanding America Study}, and \textit{Daily Kos} (detailed in \textit{Methods}). 

We present a case study on the US presidential elections between 1968--2016 in order to showcase the superior performance of our TA method. For each election year, we compare the poll estimates obtained with our PTA and ETA methods, to the statistical methods of survey averaging (SA), cumulative vote counting (CC) (see \textit{Methods} for details), and the best pollster estimations at the respective time.

Given the nature of the US presidential election system, we validate TA on a three candidate system. We refer to these as the Democratic (D), Republican (R) and "other" (O) candidate. 
The ground truth for forecasting validation represents the actual poll results from each respective election.


\section{Materials and methods}

\subsection{US Presidential Elections datasets}
Data were aggregated from \textit{Real Clear Politics} (2012a, 2016a), \textit{Understanding America Study} (2016b), and \textit{Daily Kos} (1968--2008, 2012b). Table \ref{tab:poll-datasets} provides information on all 15 datasets alongside the US presidential election results for the Democratic, Republican and "other" candidates between 1968--2016. These values are used as ground truth for measuring the performance of our TA method. Extended information on the data is found in \textit{Appendix D}.

\begin{table}[!htb]
\protect\caption{\label{tab:poll-datasets}US presidential election results between 1968--2016 expressed as percentages, the winning political party (D or R), number of polls in dataset, and pre-election duration (in days).}
\center
\begin{tabular}{l|rrrrrr}
Dataset & D(\%) & R(\%) & O(\%) & Winner & Surveys & Duration\\
\hline 
1968 & 42.7 & 43.4 & 13.9 & R & 12 & 140\\
1972 & 37.5 & 60.7 & 1.8 & R & 14 & 140\\
1976 & 50.1 & 48 & 1.9 & D & 29 & 144\\
1980 & 41 & 50.7 & 8.3 & R & 25 & 139\\
1984 & 40.6 & 58.8 & 0.6 & R & 32 & 139\\
1988 & 45.6 & 53.4 & 1 & R & 47 & 138\\
1992 & 43 & 37.4 & 19.6 & D & 64 & 144\\
1996 & 49.2 & 40.7 & 10.1 & D & 75 & 147\\
2000 & 48.4 & 47.9 & 3.7 & D & 356 & 147\\
2004 & 48.3 & 50.7 & 1 & R & 150 & 146\\
2008 & 52.9 & 45.7 & 1.4 & D & 188 & 147\\
2012a & 52.9 & 45.7 & 1.4 & D & 326 & 1276\\
2012b & 51.1 & 47.2 & 1.7 & D & 114 & 147\\
2016a & 48.2 & 46.1 & 5.7 & R & 259 & 529\\
2016b & 48.2 & 46.1 & 5.7 & R & 121 & 120\\
\hline 
\end{tabular}
\end{table}

\subsection{Alternative poll estimation methods} 

Cumulative vote counting (CC) is applied by summing up all votes expressed by the polls $p_i^*(t)$ for each candidate $i$ over the total polling period $[0, d)$. Note that for CC we do not use the normalized value $p_i(t)$, but the absolute number of votes expressed in each poll $p_i^*(t)$. Consequently, we define a cumulative momentum $cM_i(t)$ which is updates as:

\begin{equation}
cM_i(t)=
\begin{cases}
cM_i(t-1) + p_i^*(t), & \text{if }\;\,p_i(t)>0\\
cM_i(t-1), & \text{if }\;p_i(t)=0
\end{cases}
\label{eq:cc}
\end{equation}

At the end of the polling period ($t=d$), each cumulative momentum $cM_i(t)$ will store the total number of votes expressed for each candidate $i$. At any time, we can infer the current opinion towards a candidate $c\Omega_i(t)$ by normalizing each momentum:

\begin{equation}
c\Omega_i(t) = cM_i(t) / \sum_{j} cM_j(t)
\label{eq:model-omega-cumulative}
\end{equation}

Survey averaging (SA) is applied by averaging the normalized poll results over the entire pre-election period. In order to express opinion $s\Omega_i(t)$ after $t$ elapsed days, we use the normalized poll vector $p_i(t)$ directly:

\begin{equation}
s\Omega_i(t) = \frac{\sum_{0 \leq k \leq t} p_i(k)}{|\{p_i(k) | 0 \leq k \leq t\}|}
\label{eq:survey-averaging}
\end{equation}

In Equation \ref{eq:survey-averaging} we obtain the current poll at time $t$ (expressed as number of days) by summing up all normalized votes for the period $[0,t]$ and divide by the number (\textit{cardinal}) of polls in that same period.

The main distinction between CC and SA is that the first method uses the absolute number of votes for each candidate, whereas SA uses the normalized poll values.

\subsection{Determining near-optimal amplitude and damping factors}

The poll forecasting performance of the TA model depends on choosing the right $\alpha$ and $\beta$ factors. As such, we simulate forecasting on all datasets with the following values $\alpha_i \in \{0.1,0.5,1.0 \}$ for amplitude, respectively $\beta_i \in \{0.05, 0.01, 0.15, 0.2, 0.25, 0.3, 0.4 ... 0.9, 1.0, 1.25, 1.5 \}$ for damping.

For each forecasting simulation, we measure the total error $\varepsilon$ expressed as percentage offset from the final election results. Based on the real results under the form $\overline{\Omega}_{year}^{candidate}$ (given in Table \ref{tab:poll-datasets}), we define the relative estimation error $\varepsilon$ as the sum of positive differences between the estimation error for all three candidates $c \in \{D, R, O\}$:

\begin{equation}
\varepsilon_{year}(\alpha_i,\beta_i) = \sum_{c \in \{D, R, O\}} |\overline{\Omega}_{year}^{c} - \Omega_{year}^{c}(\alpha_i,\beta_i)|
\label{eq:ta-error}
\end{equation}

\begin{figure}[hbt]
\centering
\includegraphics[width=1\linewidth]{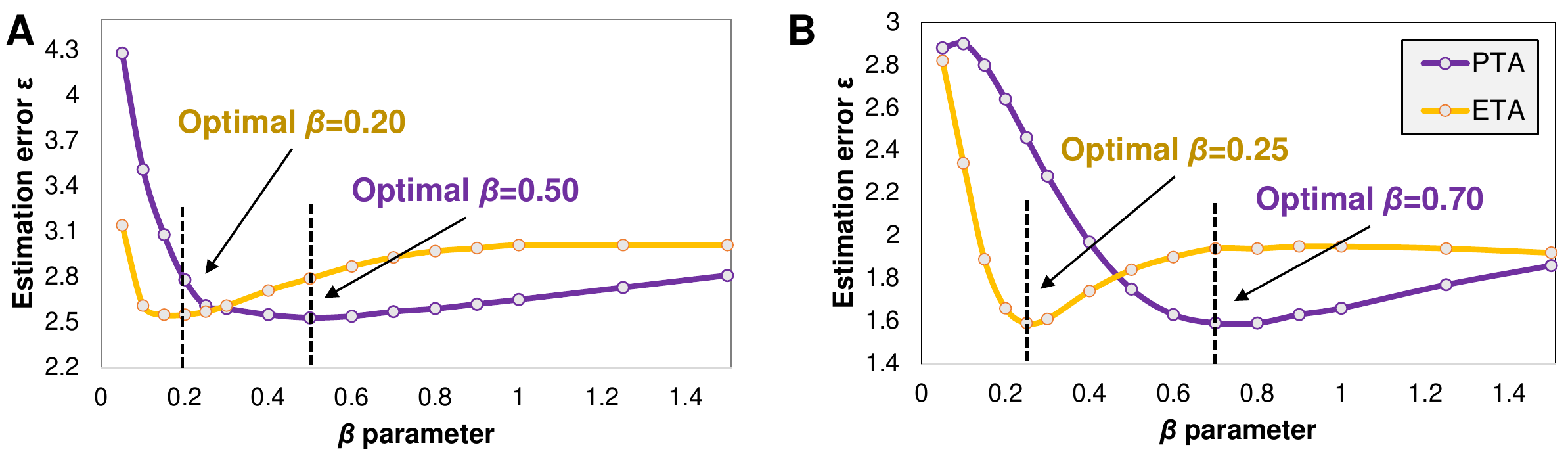}
\caption{Total poll estimation error $\varepsilon$ (D+R+O) for the PTA (violet) and ETA (yellow) methods. A smaller $\varepsilon$ is better, and we highlight the corresponding $\beta$ values with vertical dotted lines.}
\label{fig:optimal-beta}
\end{figure}

By varying $\alpha_i$ only the scale of the momentums $M_i(t)$ changes. Hence, since we normalize all $M_i(t)$ in order to obtain the poll results $\Omega_i(t)$, the value of the amplitude becomes irrelevant. On the other hand, we notice that the impact of $\beta_i$ is significant. As an example, we display in Figure \ref{fig:optimal-beta} the evolution of the total error $\varepsilon$ for the 2012 and 2016 elections, by employing PTA (violet) and ETA (yellow). Striving for a minimal error, our results suggest the following ideal $\beta_i$ values: 
$\varepsilon_{2012a}^{PTA}(\alpha=1,\beta=0.5)=2.53$, $\varepsilon_{2012a}^{ETA}(\alpha=1,\beta=0.2)=2.55$, $\varepsilon_{2016a}^{PTA}(\alpha=1,\beta=0.7)=1.59$, and $\varepsilon_{2016a}^{ETA}(\alpha=1,\beta=0.25)=1.59$.
 
The same process of finding an ideal combination ($\alpha, \beta$) can be repeated for each dataset. However, choosing an optimal damping factor \textit{during} a real-time pre-election period is not plausible because we cannot compute $\varepsilon$ without the final, real results. Therefore, we compute a pseudo-ideal $\beta$ factor from the average of the best $\beta$'s measured over all past datasets, in the perspective that it should be used for present and future predictions. As such, all forecasting simulations presented in this paper are based on the same data-derived damping factor $\beta=1.1$ for PTA, and $\beta=0.78$ for ETA.

\section{Results}

A good electoral forecasting method should have two qualities: (i) it should produce poll estimations for all candidates that are -- overall -- as close as possible to the real election results, and (ii) it should correctly determine the winner of the election.

We quantify the first property by the total estimation error $\varepsilon$ given as the sum of positive differences between the estimation error for all candidates (see Equation \ref{eq:ta-error}). A smaller $\varepsilon$ translates into a more performant forecasting method.
Table \ref{tab:poll-performance} displays $\varepsilon$ between each forecasting method and the actual election results.
Averaged over all datasets, we measure a forecasting error of 3.64 points for the best pollster, 7.31 for CC and 7.49 for SA. The errors of TA are only 3.28 points for PTA and 2.87 points for ETA. These results translate into a superior performance of TA over all other state of the art methods. We obtain a 0.364 point improvement for PTA, respectively 0.774 for ETA compared to the best pollster estimations.

\begin{table}[!htb]
\caption{Estimation error $\varepsilon$ of forecasting methods during the 1968-2016 presidential elections, alongside the best predictor at that time (name of pollster specified in parenthesis).}
\label{tab:poll-performance}
\centering
\begin{tabular}{l|rrrrl}
Dataset & CC & SA & PTA & ETA & Best\\
\hline
1968 & 9.82 & 9.78 & 4.15 & 4.11 & 4.1 (Gallup)\\
1972 & 5.63 & 5.64 & 4.02 & 1.04 & 1.8 (Gallup)\\
1976 & 9.22 & 8.97 & 1.56 & 0.53 & 0.9 (Roper)\\
1980 & 12.35 & 15.18 & 7.02 & 6.71 & 6.7 (Gallup)\\
1984 & 7.17 & 7.66 & 3.4 & 2.42 & 2.4 (Gallup)\\
1988 & 8.32 & 8.1 & 7.32 & 7.33 & 11.0 (CBS/NYT)\\
1992 & 6.52 & 5.18 & 1.06 & 1.68 & 3.4 (CBS/NYT)\\
1996 & 6.95 & 6.75 & 3.04 & 2.64 & 3.1 (Gallup)\\
2000 & 6.43 & 8.13 & 2.54 & 2.35 & 2.4 (IBD/CSM)\\
2004 & 7.18 & 6.54 & 1.73 & 1.74 & 2.7 (RCP)\\
2008 & 7.59 & 6.67 & 2.47 & 1.81 & 2.0 (RCP)\\
2012a & 7.3 & 6.88 & 2.53 & 2.55 & 3.2 (RCP)\\
2012b & 5.28 & 4.95 & 3.16 & 2.95 & 3.2 (RCP)\\
2016a & 3.91 & 5.72 & 1.59 & 1.59 & 3.9 (RCP)\\
2016b & 5.99 & 6.17 & 3.64 & 3.63 & 3.9 (RCP)\\
\hline
\end{tabular}
\end{table}

The second property is quantified by the offset from the real difference between the Democratic and Republican candidates (D-R), expressed in percentage points. Table \ref{tab:poll-d-r} represents the real (D-R) differences for all election years, followed by the relative offsets of each forecasting method from this difference. For example, in 1968, (D-R)=-0.7 and the offset of PTA is -1.43 (we can infer that PTA estimated the (D-R) difference at $-0.7+(-1.43)=-2.13$ points).
By averaging all forecasting offsets (in positive value) over all datasets, we measure a total (D-R) offset of 1.95 points for the best pollster, 4.07 for CC and 3.59 for SA. TA outperforms again the other methods by scoring an average offset of 1.74 for PTA and 1.91 for ETA.

\begin{table}[!htb]
\caption{Column (D-R) represents the absolute percentage difference between the Democratic and Republican candidates after election. All subsequent columns represent the relative offset of the forecasting methods from the (D-R) difference. A smaller offset means a closer estimation of the winning party.}
\label{tab:poll-d-r}
\centering
\begin{tabular}{l|rrrrrr}
Dataset & (D-R) & CC & SA & PTA & ETA & Best\\
\hline
1968 & -0.7 & -6.12 & -5.8 & -1.43 & -1.31 & -1.3\\
1972 & -23.2 & -3.51 & -3.52 & 0.24 & -1.04 & -0.8\\
1976 & 2.1 & 8.46 & 7.93 & 0.12 & 0.53 & 0.9\\
1980 & -9.7 & 7.15 & 3.42 & 7.02 & 6.71 & 6.7\\
1984 & -18.2 & 3.43 & 3.32 & 1.16 & 1.2 & 1.2\\
1988 & -7.8 & 6.14 & 6.54 & -0.14 & -1.31 & -0.2\\
1992 & 5.6 & -6.52 & -5.18 & -1.06 & 1.24 & 3.4\\
1996 & 8.5 & 6.95 & 6.75 & 3.04 & 2.64 & 2.5\\
2000 & 0.5 & -2.47 & -2.49 & -0.38 & 1.25 & -2.4\\
2004 & -2.4 & 1.12 & 0.58 & 1.73 & 1.56 & 0.9\\
2008 & 7.2 & -3.07 & -2.17 & 0.51 & 0.83 & 0.4\\
2012a & 3.9 & -1.36 & -1.2 & -2.53 & -2.55 & -3.2\\
2012b & 3.9 & -1.62 & -1.73 & -3.16 & -2.95 & -3.2\\
2016a & 2.1 & 2.89 & 2.94 & 1.59 & 1.59 & 1.1\\
2016b & 2.1 & -0.27 & -0.25 & 2 & 1.89 & 1.1\\\hline
\textit{Average} & - & 4.07 & 3.59 & 1.74 & 1.91 & 1.95\\
\hline
\end{tabular}
\end{table}

Additionally, we investigate how many out of the 13 elections (1968--2016) are correctly predicted in terms of picking the right winning party. As such, the reference statistical methods are the least performant, with CC predicting only 10 out of 13 presidents, and SA predicting 11 out of 13. The best pollsters manage to predict 11 out of 13, while PTA and ETA predict 12 out of 13 winners. No method was able to predict the correct 2016 winner. We did find one pollster which correctly predicted the winner of those elections, but did so by a much greater error in terms of popular vote. Supporting results for Tables \ref{tab:poll-performance}, \ref{tab:poll-d-r} are provided in \textit{Appendix E}.

\begin{figure}[hbt]
\centering
\includegraphics[width=0.75\linewidth]{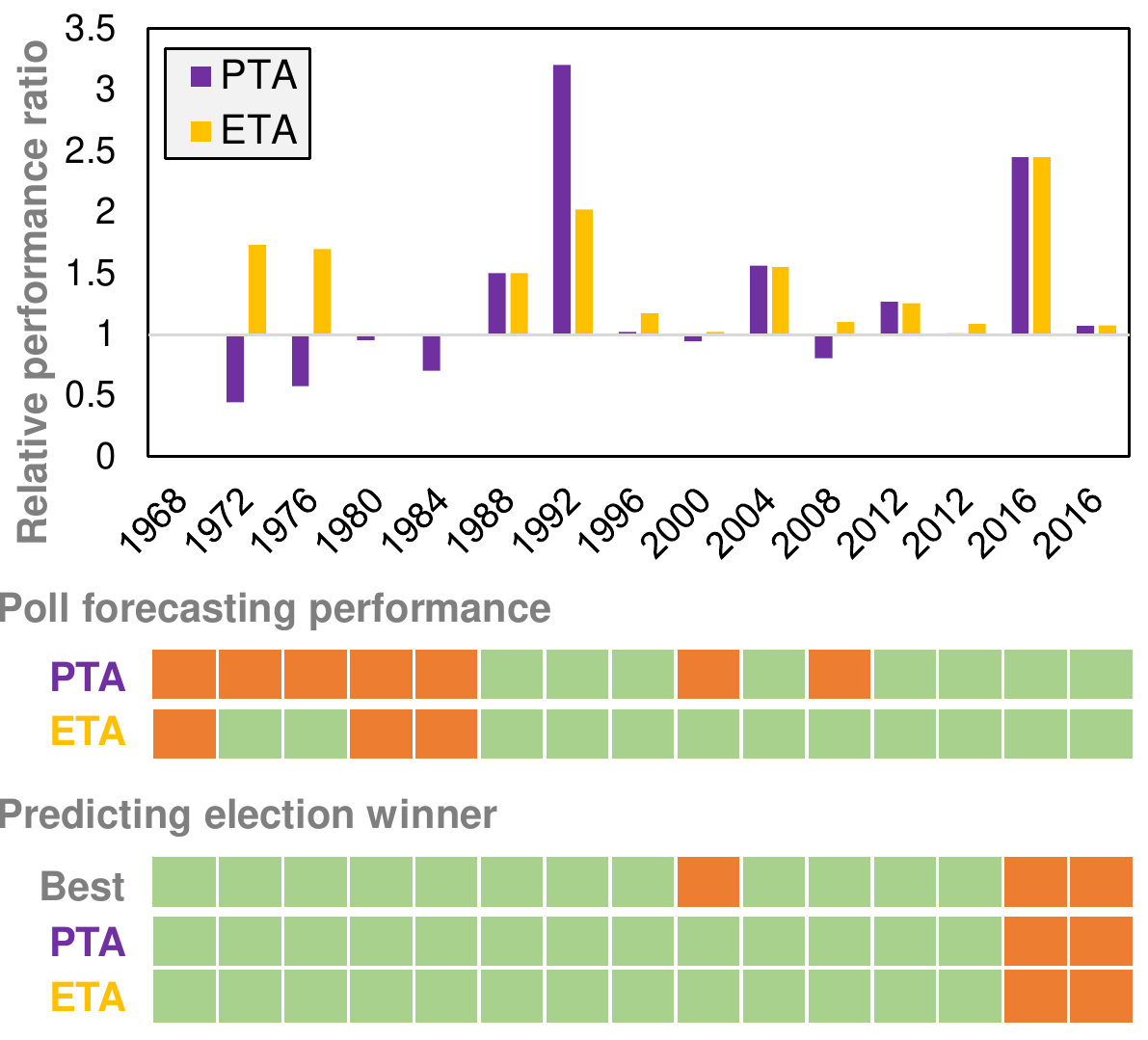}
\caption{Ratio between the best pollster prediction error and the PTA (violet), respectively, ETA (yellow) prediction error. Values above 1 translate into higher prediction performance of our TA methods. In the lower panel we highlight the cases when TA outperforms the best pollster (green, or red otherwise), and the election years in which any of the three predictors manage to forecast the correct election winner.}
\label{fig:ta-performance}
\end{figure}

Furthermore, we highlight in Figure \ref{fig:ta-performance} the superior estimation performance of TA compared to the state of the art. We underline the fact that our TA methods outperform the best pollster, in terms of estimation error $\varepsilon$, on 8 out of 15 datasets (PTA), respectively 12 out of 15 datasets (ETA). Overall, TA outperforms the competing predictors in 10 out of 13 election years (77\%). The upper panel in Figure \ref{fig:ta-performance} graphically represents the ratio between the best pollster prediction error and the TA method prediction error (\textit{i.e.}, $\varepsilon_{Best}/\varepsilon_{TA}$). As such, values (represented as columns) over 1.0 mean a higher performance for our TA methods. The lower panel in Figure \ref{fig:ta-performance} classifies the cases when our PTA or ETA methods outperform the best pollster prediction (\textit{i.e.}, with green, otherwise red), for each election dataset. Also, we represent the cases when any of the three methods compared manage to forecast the correct winner of the elections. The only notable difference is that the best pollster does not succeed to forecast the winning party in the 2000 elections. 

\subsection{Real time feasibility analysis} 

We extend our analysis by exploring the feasibility of our theoretical framework in the context of application during a real time pre-election period. Thus, we use the 2016a (RCP) dataset and compare the prediction errors $\varepsilon$ at different points in time. The dataset consists of 259 polls over a 529 day period prior to the elections. We choose arbitrarily to measure the predictions and corresponding $\varepsilon$ at $t=\{100,250,400,500,529\}$. Table \ref{tab:polls-real-time} represents the estimation errors $\varepsilon$ for all five forecasting methods (here the best pollster is RCP), and the improvement ratio is $\varepsilon_{RCP}/\varepsilon_{ETA}$. Values greater than 1 mean higher performance for ETA. Detailed experimental results are given in \textit{Appendix B}.

\begin{table}[!htb]
\caption{Estimation error $\varepsilon$ of forecasting methods during the 2016 presidential pre-election period at different moments in time (here, t=0: 2015/05/26 and t=529: 2016/11/08). In the last column, the ratio between the RCP estimation error and the ETA error.}
\label{tab:polls-real-time}
\centering
\begin{tabular}{l|rrrrrr}
$t$ & CC & SA & PTA & ETA & RCP & Ratio\\
\hline
100 & 9.88 & 10.41 & 3.82 & 3.69 & 6.7 & 1.81\\
250 & 5.16 & 6.94 & 6.95 & 7.32 & 2.8 & 0.38\\
400 & 6.95 & 7.82 & 6.6 & 6.55 & 10.2 & 1.55\\
500 & 4.83 & 6.72 & 4.58 & 4.57 & 4.8 & 1.05\\
529 & 3.91 & 5.72 & 1.59 & 1.59 & 3.9 & 2.45\\
\hline
\end{tabular}
\end{table}

We find that the prediction accuracy of the statistical methods (CC, SA) depends mainly on the amount of data, as their forecasts slowly converge towards the real results. Conversely, the other methods do not depend on the increasing amount of data (as we get closer to the election day), but rather on the volatility of the socio-political context. Namely, RCP has the highest fluctuations, registering jumps from a low $\varepsilon=2.8$ (February 2016) to a very high $\varepsilon=10.2$ (July 2016), then falling back to $\varepsilon \approx 4-5$ (October 2016). Our TA methods register more stability than RCP, and are not influenced by the same social volatility we measured in RCP. We note that the Democratic candidate gathered increasing popularity until March 2016, so that TA reflects this by giving her a higher virtual chance of winning. Nevertheless, as the popularity of the Republican candidate rapidly increased, during mid-spring and mid-summer 2016, our forecasting becomes better leaning towards a balanced outcome that is closer to the final registered popular vote.

We have found that, unlike the best pollsters, which rely on MRP corroborated with social, economical and political trends, our TA method improves its forecasting based solely on the time-aware convergence of public opinion, which can be considered of significant estimation prowess \cite{lewis1999voters,becker2017network,sjoberg2009all}.  

\section{Discussion}

Our study differs in several respects from previous work on election forecasting. In comparison to basic statistical approaches, like CC and SA, our TA needs additional temporal information on each pre-election poll (\textit{i.e.}, date when a poll was made public). Unlike simple averaging of the information, we feed the survey data to our simulation framework which is highly influenced by the temporal aspect. Both PTA and ETA methods model opinion momentum, in time, as a function which bounces up when opinion is injected, and dampens down otherwise. This process resembles the way a capacitor charges (\textit{i.e.}, opinion being injected) and discharges slowly (\textit{i.e.}, relaxation state, no opinion injected). Compared to the state of the art methods, like MRP \cite{christensen2008predicting,kiewiet2018predicting,hummel2014fundamental}, our TA does not need any demographic, economic, or political information related to the context of the election. This distinction represents a significant advantage for TA over MRP since our method may be applied, given enough reliable public polls, in any political region of the world. Similar to the case study in this paper, we did not consider any additional information about the USA during the 1968--2016 period.

In essence, the TA model is aimed at improving the prediction of the \textit{popular vote}. Nevertheless, we find studies especially tailored to systems like the US, which are based on the college system \cite{hummel2014fundamental,kiewiet2018predicting}, and, conversely, tailored to systems utilizing a direct popular vote, like France \cite{nadeau2010electoral}. The work of \cite{hummel2014fundamental,kiewiet2018predicting} manages to forecast US presidential, senatorial, and gubernatorial elections at the state level by incorporating state level demographics to better predict the college vote.
However, we have developed the TA forecasting model to be usable outside any political context, as long as there is sufficient and reliable pre-election poll data. This choice may give it an apparent disadvantage in the US system, but as our case study was intended to show, in practice TA still yields superior performance. Moreover, where other models may need specific tuning to be used in other countries of the world, TA will work without the need for customization.

In this study we start from the premises that the opinion injected in social networks, stemming from publicly accredited opinion polls, has a very high media coverage. To this end, recent studies on how US adults keep themselves informed about political candidates and issues, show that TV (news) occupies the leading spot with 73\%, followed by 45\% for news websites/apps, 24\% for newspapers, and 21\% specifically for social media. These statistics are in favor of our premises since opinion injection from pollsters is practically done through all the enumerated media types \cite{graber2017mass}.
Furthermore, in terms of polling reliability, current media types are diverse, but their combined coverage remains high, including in the electoral context, and polling accuracy remains reliable \cite{groves2006nonresponse}.

Finally, as an explanation to why TA outperforms more complex data-driven methods used by pollsters (\textit{e.g.}, MRP), we notice that the forecasts of TA for the "other" (O) candidates are lower, and implicitly closer, to the real results. Averaged over all datasets, the pre-election surveys predict that 11.61\% will vote for the O candidate; TA predicts 6.94\%, and the best pollster predicts 7.57\%. However, following each election, we compute the real average percentages for the O candidate at only 5.08\%. This means that, even though the forecasts for the D and R candidates may be realistic, the public opinion polls are unable to distribute a difference of $\approx 6.5\%$ of remaining votes. On the other hand, the best pollsters are unable to distribute $\approx 2.5\%$ of overall votes, and TA only $\approx 1.8\%$ of votes.
This observation does not mean that TA simply overestimates the percentages for the two main candidates; it means that TA is able to distribute the votes for the O candidate more realistically based on the dynamics of expressing opinion just before the end of the pre-election period, which usually sees an abrupt drop of $\approx 30\%$ in popularity for O (see an extended statistical analysis in \textit{Appendix C}. 
Of course, these performance gains of TA can be further analyzed in future research, and supported by both social psychology or political science assumptions.

\subsection{Limitations of the model}

Our TA model brings some limitations along, which we further discuss. 
For instance, we consider social media as an ubiquitous diffusion mechanism, but there are also, so called, non-users. We added this form of simplification to our model due to difficulty in acquiring data for offline users, and due to the reliability of that data. 
Official statistics approximate that 3/4 of the US population are engaged in social media. Even in this case, we argue that our model's simplification remains robust, as a study on political attitudes concludes that no statistically significant differences arise between social media users and non-users on political attention, values or political behavior \cite{mellon2017twitter}

Another realistic simplification in our model allows us to consider the electoral system relatively hard to shape from the outside, so that we do not have to account for data beyond our reach (\textit{i.e.}, external influences). The liberal democracy index was developed to measure the robustness of a political system, and, according to a study by the Swedish V-Dem institute, the USA scores 0.75 (out of 1) and lies within the top 20\% liberal nations \cite{coppedge2019v}. As such, we can consider the studied US electoral system as robust.

Existing vote polarization and poll credibility are also important topics to consider in the future\cite{bernhardt2008political}, however, our electoral forecasting model was designed to be, as much as possible, unaffected by any social and political contexts, including the effects of opinion polarization.

\subsection{Conclusions}

Driven by the increases in access to data and computational power, modern election forecasting systems should, intuitively, evolve along one of two directions: a possible microscopic framework built on extensively detailed social media data, or a possible macroscopic framework employing complex data science techniques on demographics and economic indices. However, our proposed model represents a trade-off between both micro and macro worlds, and the result is a simple, intuitive and robust methodology which can be applied on any pre-election data with temporal information. We argue that this simplification is effective since social influence often pertains to the knowledge of crowds \cite{becker2017network}. In other words, the aggregated judgment of many individuals (macro-scale) can be more accurate than the judgments of individual experts (micro-scale) \cite{sjoberg2009all}. This effect is significantly strengthened when applied on larger population sizes \cite{becker2017network}.

Despite the apparently simplified assumptions behind TA, revolving around the idea that we can apply a microscopic opinion interaction model to predict macroscopic behavior, our results pinpoint to the fact that time-awareness is more significant in poll forecasting than previously considered.
In our case study, TA outperforms state of the art election forecasting methods in 10 out of the 13 presidential elections. TA accumulates an average forecasting error of 2.87--3.28 points, while statistical methods accumulate 7.48 points error, and the best pollster estimations accumulate 3.64 points. 
This translates into a roughly 30\% prediction improvement for our method, in terms of forecasting accuracy of the popular vote. 

Moreover, analyzing the methods of reputable institutions in the US, like the \textit{Huffington Post}, \textit{Real Clear Politics}, or \textit{Five Thirty Eight}, we have not seen any temporal attenuation method that is similar to the one proposed in this paper.
Other statistical, or data science approaches (\textit{e.g.}, MRP) rely on specific social, economical, and political contexts to improve and tune their predictions. Conversely our TA does not require socio-economical contextual information, and we believe that this independence translates into an advantage. It will probably never be possible to create the perfect forecasting system, due to the complexity of elections, but our TA represents a novel and distinguishable scientific proposal with proven high performance.


\section*{Conflicts of interest}
The authors declare that there are no conflicts of interest.

\section*{Acknowledgments}

This study relies partially on data from survey(s) administered by the Understanding America Study, which is maintained by the Center for Economic and Social Research (CESR) at the University of Southern California. The content of this paper is solely the responsibility of the authors and does not necessarily represent the official views of USC or UAS. 

\bibliography{article}

\end{document}